\documentclass[12pt,preprint]{aastex}
\usepackage{natbib}
\usepackage{graphicx}

\shorttitle{Asymmetries in coronal spectral lines in Moss}
\shortauthors{Tripathi and Klimchuk}

\begin{document}

\title{Asymmetries in Coronal Spectral lines and Emission Measure Distribution}
\author{Durgesh Tripathi}
\affil{Inter-University Centre for Astronomy and Astrophysics, Post Bag - 4, Ganeshkhind,
Pune 411007, India}
\and
\author{James A. Klimchuk}
\affil{NASA Goddard Space Flight Center, Greenbelt, MD 20771, USA}

\begin{abstract}
It has previously been argued that 1. spicules do not provide enough
pre-heated plasma to fill the corona, and 2. even if they did,
additional heating would be required to keep the plasma hot as it
expands upward. We here address the question of whether spicules
play an important role by injecting plasma at cooler temperatures
($< 2$ MK), which then gets heated to coronal values at higher
altitudes. We measure red-blue asymmetries in line profiles formed
over a wide range of temperatures in the bright moss areas of two
active regions. We derive emission measure distributions from the
excess wing emission. We find that the asymmetries and emission
measures are small and conclude that spicules do not inject an
important (dominant) mass flux into the cores of active regions at
temperatures $> 0.6$ MK ($\log T > 5.8$). These conclusions apply
not only to spicules, but to any process that suddenly heats and
accelerates chromospheric plasma (e.g., a chromospheric nanoflare).
The traditional picture of coronal heating and chromospheric
evaporation appears to remain the most likely explanation of the
active region corona.
\end{abstract}


\section{Introduction}

There has been great interest recently in fast ($\sim 100$ km
s$^{-1}$) upflows revealed by asymmetries in the profiles of hot
spectral lines. These asymmetries take the form of small
enhancements in the blue wings of the lines. They have been
measured by numerous authors using a variety of techniques,
including double Gaussian fits and intensity differences between the
red and blue sides of the profile
\citep{HarW:08,DeMH:09,MacD:09,BryYD:10,MarDH:11,TiaMD:11,
TiaMX:12,Dos:12,BroW:12}.
The asymmetries tend to be very subtle. They are best seen in faint
areas at the peripheries of active regions, where the intensity of
the secondary (upflow) component can exceed 20\% of the intensity of
the main (``rest'') component. The intensity ratio is $< 5\%$ over
most of the Sun.

It has been suggested that these upflows are associated with type~II
spicules \citep{DeMH:09}. Spicules originate in the chromosphere and
are mostly cold, but limb observations from the Atmospheric Imaging
Assembly \citep[AIA;][]{LemEtal:12} on the Solar Dynamics
Observatory (SDO) show that the tips of at least some type~II
spicules are heated to coronal temperatures as they are ejected
\citep{DeMC:11}. The speeds of the proper motions seen at the limb
are comparable to the Doppler shifts of the blue wing enhancements
seen on the disk ($\sim 100$ km s$^{-1}$). It was proposed that
this hot spicule plasma may explain much or even most of the
material that we observe in the corona. This would be a major shift
from the conventional picture in which heating occurs in the corona
itself, i.e., either the coronal plasma is maintained at high
temperatures by quasi-steady heating, or else the plasma undergoes
repeated cycles of impulsive heating (nanoflares) and evaporation,
followed by slower cooling and draining.

\citet{Kli:12} recently examined the role of type~II spicules in
supplying the corona with hot plasma. He started with the hypothesis
that all coronal plasma comes from spicules and examined the
observational consequences. One test concerns the ratio of emission
measure in the upflow to the emission measure in the downflow.
\cite{Kli:12} found that, if the hypothesis of a spicule-dominated corona
is correct, this ratio must exceed 3 in active regions, 1 in the
quiet Sun, and 0.7 in coronal holes.  He went on to argue that the
temperatures of the upflow and downflow should be similar and
therefore the emission measure ratio would approximately equal the
ratio of blue wing to line core intensities in a typical coronal
spectral line from species like \ion{Fe}{14}. Since the observed ratios
are much smaller than predicted, Klimchuk concluded that spicules
can contribute at most a small fraction of the corona's hot plasma.
In active regions, the discrepancy is two orders of magnitude. A
second test involves the ratio of emission measures in the corona
and lower transition region.  Again the predicted and observed
values disagree by two orders of magnitude. A third test was
proposed involving the ratio of densities in the fast upflow and
slow downflow. \citet{PatKY:13} recently found that the discrepancy
is here, too, approximately two orders of magnitude.  It appears
that the quantity of hot plasma ejected into the corona by type~II
spicules---indeed by any mechanism that suddenly heats and
accelerates chromospheric plasma---is far less than the amount of
hot plasma that exists in the corona.

\citet{Kli:12} also pointed out that, even if enough hot plasma were
ejected, additional heating would be required at higher altitudes to
compensate for the extreme adiabatic cooling that would otherwise
occur as the plasma expands upward to fill the coronal flux tube. In
the absence of heating, a 2~MK plasma at the tip of a spicule would
cool to roughly 0.1~MK from expansion alone. Strong heating must
therefore occur in the corona itself.

Even if spicules do not
directly provide enough hot plasma to explain the corona, they might
nonetheless eject large amounts of cooler plasma that is then heated
to coronal temperatures at higher altitudes.  We know that the bulk
of a type II spicule is heated to $< 0.1$~MK as it is ejected
\citep{DeMC:11}. However, this material falls back to the surface
after reaching a maximum height of only $10^4$ km, so it does not
contribute to the coronal mass. Is it possible that a significant
amount of material is heated to higher temperatures during the
ejection, though still below 2~MK, and then later heated all the way
to coronal values? This is the question that we address in this
paper. Our approach is to determine the emission measure
distribution of the ejected plasma by examining the excess blue wing
emission of multiple spectral lines covering a range of
temperatures. We note in passing that the spicule material that is
only weakly heated to $< 0.1$~MK may explain the very bright
emission from the lower transition region, which conventional
coronal heating models cannot \citep[][although see
\cite{AntN:86} for an alternative explanation in mixed polarity
quiet Sun regions]{Kli:12}.

\citet{BroW:12} have recently published results on emission measure
distribution of excess blue wing emission observed in a faint
``outflow'' region at the periphery of an active region.  As already
mentioned, these are the places where line asymmetries are greatest.
They are called outflow regions not because of the asymmetries, but
because the line core is significantly Doppler shifted to the blue.
It has been proposed that they may be an important source of slow
solar wind \citep[e.g.,][see also \cite{DelAK:11} for an alternative explanation]{SakKN:07, HarSM:08}. Using double Gaussian
fits, \cite{BroW:12} found that
the EM of the secondary component peaks at 1.4{--}2.0~MK and decreases
rapidly with temperature below 1~MK. The ratio of the EM at 0.6~MK
($\log T=5.8$) to the maximum EM is generally $< 0.01$. In this faint
outflow region, therefore, the EM determined from an \ion{Fe}{14} line is
a reasonable estimate of the total EM in the upflow.

Our study concerns bright moss areas within two active regions. Moss
corresponds the footpoints of unresolved loops that form the
generally hot and diffuse cores of active regions
\citep[e.g.,][]{MarKB:00,TriMYD:08, TriMD:10}. Any flows seen at
these locations would be related to the mass balance of the core
plasma.  In particular, any spicules that supply mass to the core of active regions
would occur here.  As we will show, and has been reported previously
\citep[e.g.][]{Dos:12}, line asymmetries are much smaller at these
locations and are a significant challenge to measure.  We therefore
employ a new technique called Intensity Conserving Spline
Interpolation \citep[ICSI;][]{KliPT:13}. Even with this technique, we
are only able to place upper limits on the EM of the excess wing
emission. These upper limits are nonetheless extremely useful for
evaluating the role of spicule upflows in supplying plasma to the
corona.

\section{Definition of RB-asymmetry}\label{rb}

\begin{figure}
\centering
\includegraphics[width=0.5\textwidth]{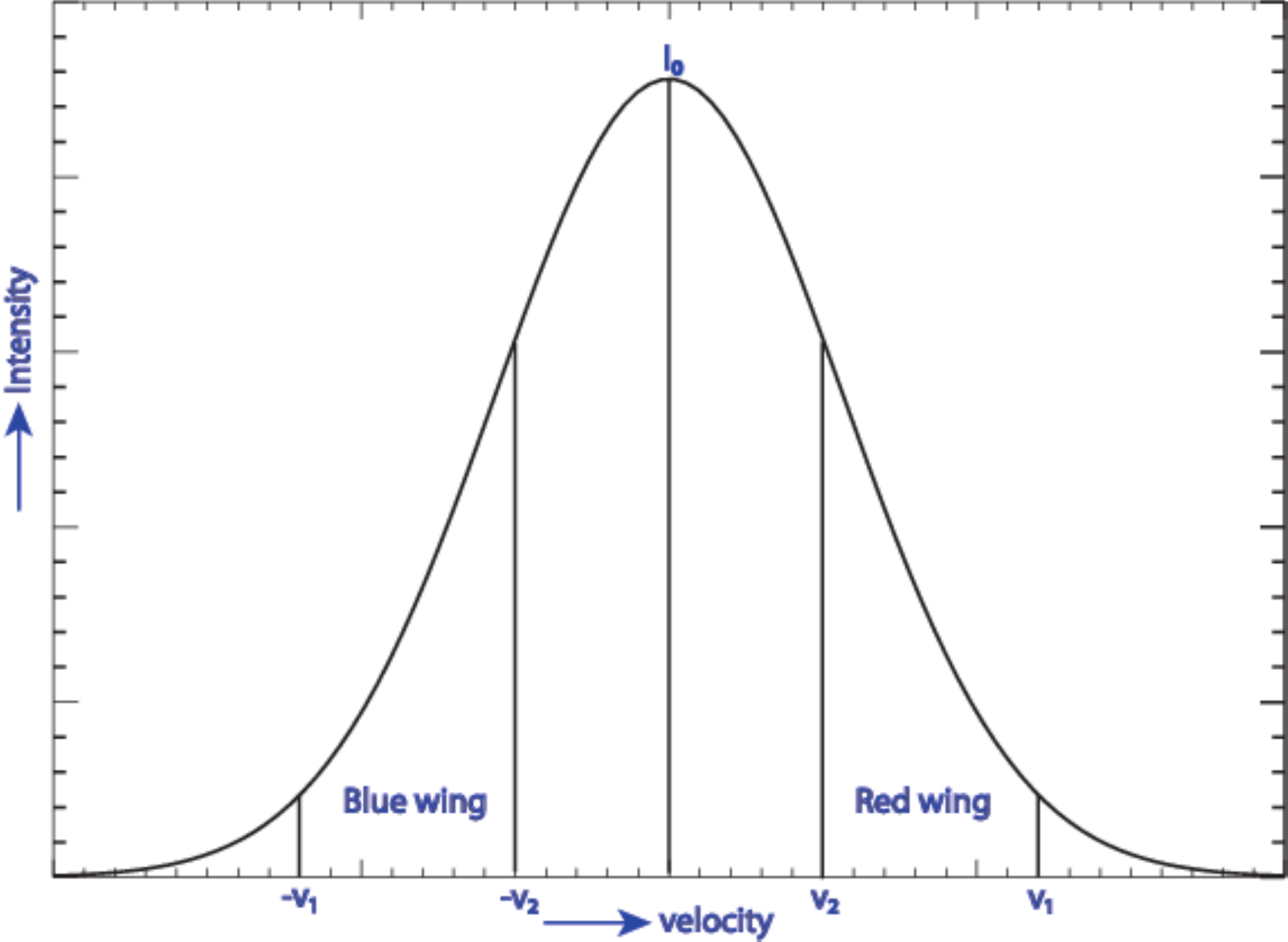}
\caption{A schematic of a line profile to demonstrate the definition
of RB asymmetry \label{gaussian_cartoon}}
\end{figure}

Following many other studies, we isolate the excess wing emission
using the so-called red-blue (RB) asymmetry.
 Figure~\ref{gaussian_cartoon} displays a schematic Gaussian to
demonstrate the RB asymmetry. We define the total emission in the
blue wing of the spectral line within the velocity range v1 and v2
as:

\begin{equation}
\begin{array}{l}
\displaystyle I_B = \int_{-v1}^{-v2} I(v)\,dv  = \sum I_i~\Delta{v}
= \Delta{v} \sum_i I_i ,\\  \label{eq:blue}
\end{array}
\end{equation}

\noindent where $\Delta~v$ is the grid size. Similarly, the total
emission in the red wing in the same velocity range towards the
positive velocity side can be written as

\begin{equation}
\begin{array}{l}
\displaystyle I_R = \int_{v2}^{v1} I(v)\,dv  = \sum I_i~\Delta{v} =
\Delta{v} \sum_i I_i . \label{eq:red}
\end{array}
\end{equation}

If we consider that the emission in the core of the spectral line is
part of a Gaussian profile given as

\begin{equation}
\displaystyle I(v) = I_0~exp\left[-\left(\frac{v}{w}\right)^2\right]
,
\end{equation}

\noindent where I$_0$ is the line center intensity, and $w$ is the
$1/e$ halfwidth, then the total emission in the Gaussian is

\begin{equation}
\displaystyle I_{core} = \sqrt{\pi}~w~I_0 .
\end{equation}

\noindent If the broadening is just because of thermal motions, then
$w$ can be written as

\begin{equation}
w = \frac{\lambda_0}{c}~\sqrt{\frac{2kT}{M}} ,
\end{equation}

\noindent where $\lambda_0$ is the centroid of the line, $T$ is the
peak formation temperature, and $M$ is the atomic mass of the ion.
$k$ is Boltzmann's constant and $c$ is the speed of light.

The RB (Red-Blue) asymmetry can then be defined as

\begin{equation}
\begin{array}{l}
\displaystyle RB = \sqrt{\frac{1}{\pi}}  ~~ \frac{\Delta~v}{w}   ~~\frac{\sum I_B - \sum I_R}{I_0}
\end{array}\label{finalequation}
\end{equation}

\noindent However, it is well known that in transition region and
coronal spectral lines non-thermal broadening tends to dominate. In
the present analysis, we have taken $w$ = 30~km~s$^{-1}$ for all the
spectral lines. We note that the most dominant line broadening is instrumental 
which is about 55~m{\AA} for the Extreme-ultraviolet Imaging Telescope (EIS) and which 
may vary over the CCD. The net broadening may vary over the field of view, but we have 
verified that the assumption of a constant 30~km~s$^{-1}$ does not affect our measurements of 
the RB asymmetry in a significant way.

\section{Observations} \label{obs}

\begin{figure}
\centering
\includegraphics[width=0.8\textwidth]{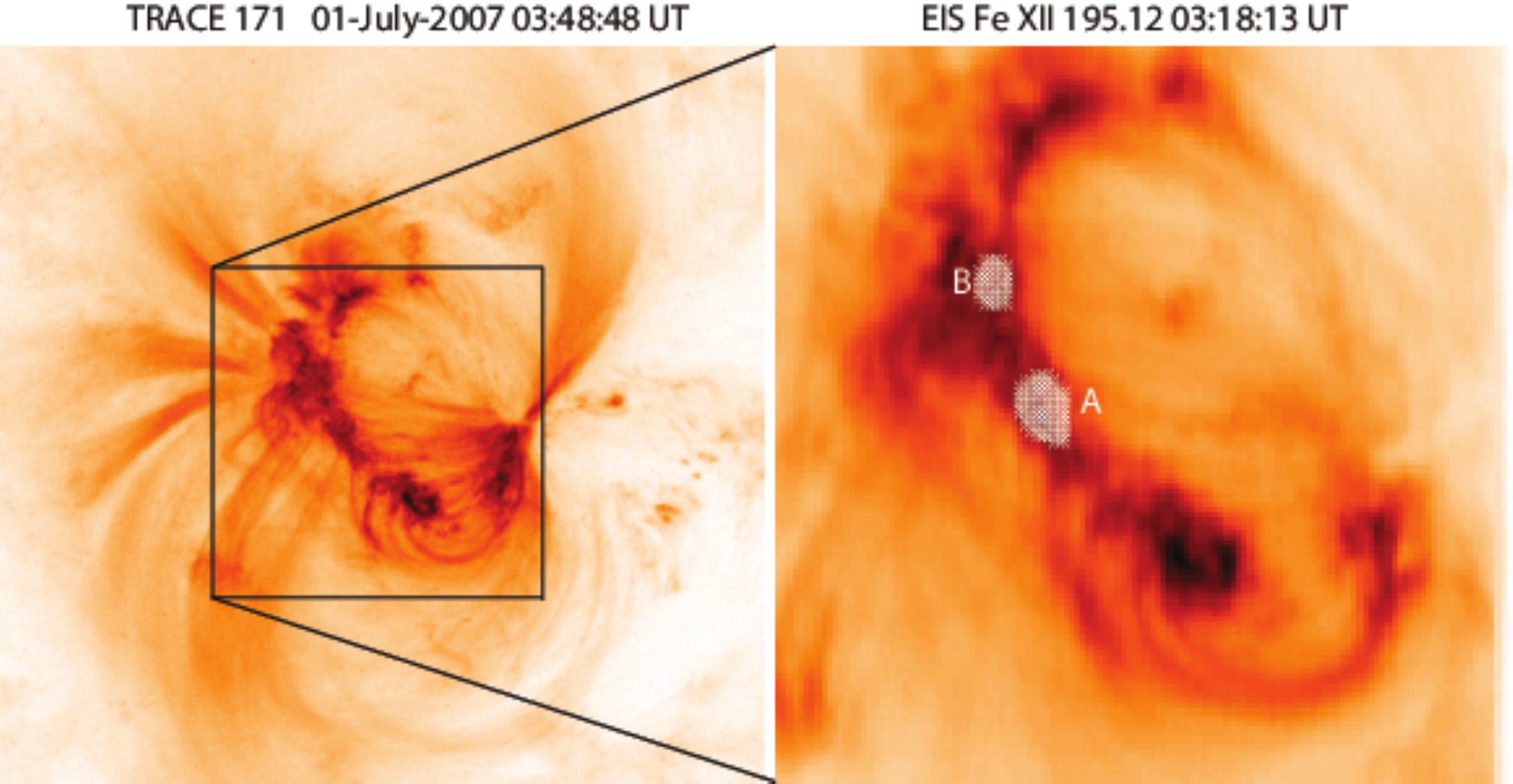}
\caption{Left panel: AR 10961 observed by TRACE on July 01, 2007. The rectangular box show the EIS FOV.
Right panel: EIS image obtained in \ion{Fe}{12} line. Regions A and B are the identified moss regions for the
study of RB asymmetry. The figure is adopted from \cite{TriMK:10}}\label{july1}
\end{figure}

In this study we have analysed observations recorded by the
Extreme-ultraviolet Imaging Spectrometer \citep[EIS;][]{CulEtal:07}
on board \textsl{Hinode}. We have used observations for two active
regions, namely \textit{AR~10961} and \textit{AR~10953}. The EIS
raster for the active region \textit{AR~10961} was recorded on 2007
July 1 at 03:18~UT and is shown in Fig.~\ref{july1}. The left panel
of the figure displays an image recorded in 171~{\AA} from the
Transition Region and Coronal Explorer \citep[TRACE;][]{HanEtal:99}.
The rectangular box on the TRACE image represents the field-of-view
(FOV) of the EIS raster, which is 128{\arcsec} by 128{\arcsec}. The
1{\arcsec} slit with an exposure time of 25~s was used. An image
obtained in \ion{Fe}{12}~195~{\AA} from the raster is shown in the
right panel of the figure. Also shown in the right panel are the two
moss regions, namely 'A' and 'B' which are chosen for the analysis.

\begin{figure}
\centering
\includegraphics[width=0.8\textwidth]{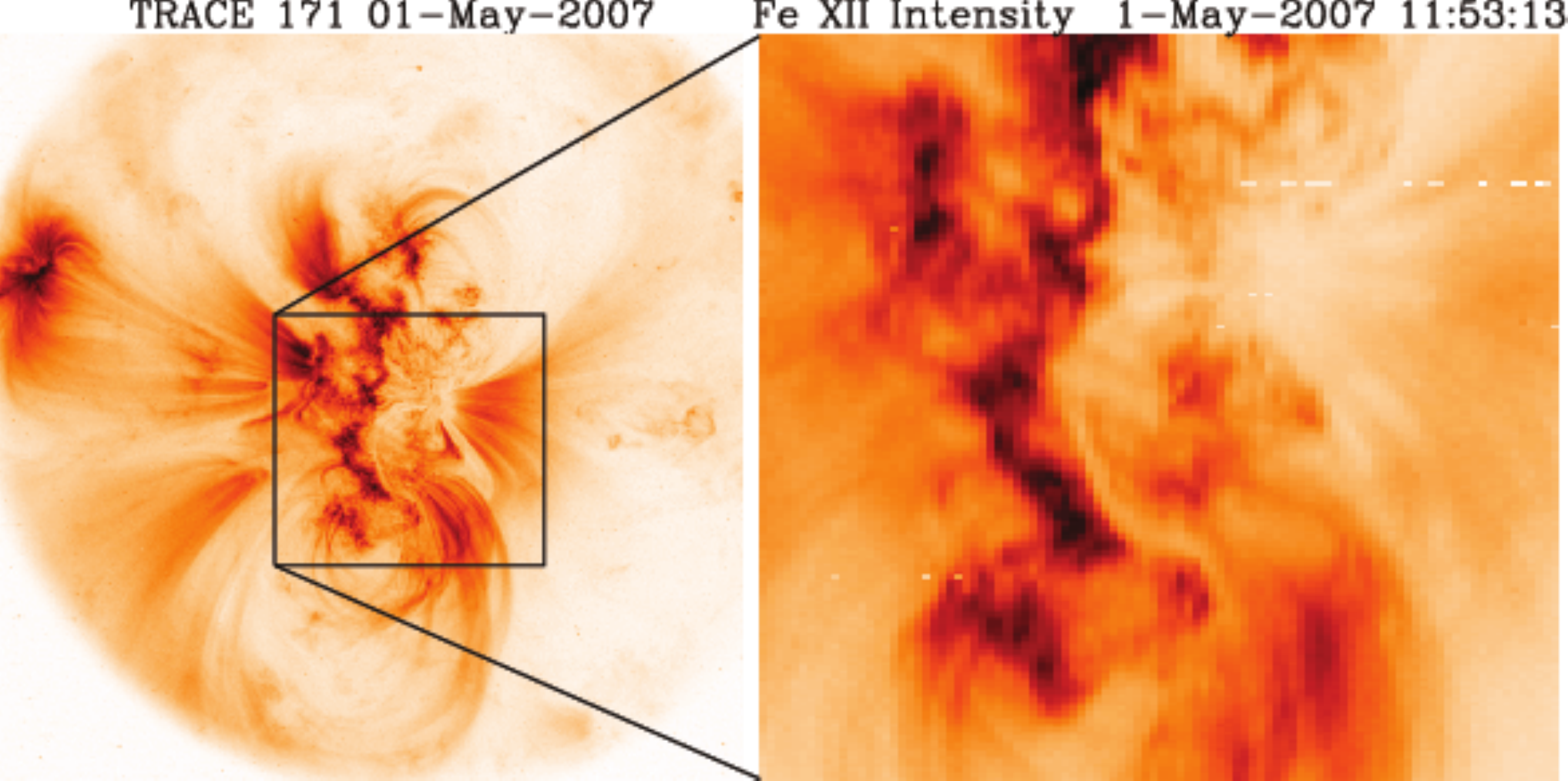}
\caption{Left panel: AR 10953 observed by TRACE on May 01, 2007. The rectangular box show the EIS FOV.
Right panel: EIS image obtained in \ion{Fe}{12} line.}\label{may1}
\end{figure}

\begin{figure}
\centering
\includegraphics[width=0.85\textwidth]{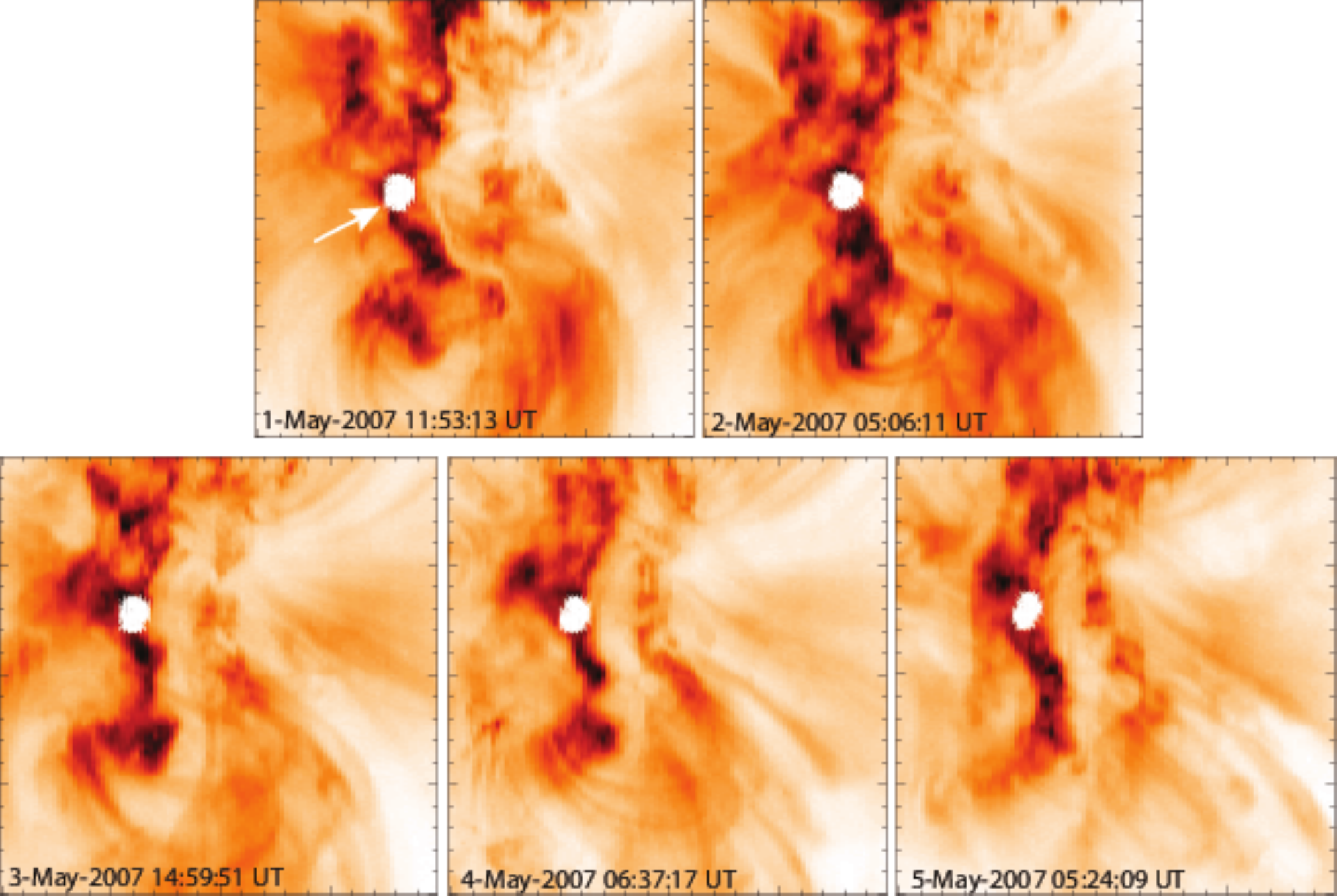}
\caption{EIS \ion{Fe}{12} images of AR 10953 observed for five consecutive days. The marked regions is identified
moss region for which the RB asymmetry is studied. \label{tracked_region}}
\end{figure}

We have also studied the center-to-limb variation of the RB
asymmetry and corresponding emission measure distribution. For this
purpose we selected \textit{AR~10953}. This active region was
observed for five consecutive days (from May 01 to May 05, 2007)
using the study sequence \textsl{CAM\_ARTB\_CDS\_A}. During this
time the active region moved from the disk center towards the
western limb of the Sun. An image of the active region recorded by
TRACE using the 171~{\AA} passband on May 01, 2007 is shown in
Fig.~\ref{may1}. The box on the TRACE image shows the EIS field of
view. A corresponding image obtained in \ion{Fe}{12}~195~{\AA} is
shown in the right panel. This study sequence takes about 20 minutes
to raster a FOV of 200{\arcsec} by 200{\arcsec} with an exposure
time of 10 s using the 2{\arcsec} slit. The moss region which was tracked is
shown in Fig.~\ref{tracked_region} by an arrow.

The rasters used here have been studied earlier for different
purposes and the details can be found in \cite{TriMYD:08, TriMD:10}
for \textit{AR~10953} and in \cite{TriMK:10, TriKM:11} for
\textit{AR~10961}. We have applied standard data processing software
provided in \textsl{solarsoft} \citep[][]{FreH:98} to each dataset.

\section{Data Analysis and Results}\label{analysis}

In order to derive the RB asymmetry, it is important to compute the line
center position (LCP) as accurately as possible. A variety of
methods have been used. \citet{DeMH:09} fitted a single Gaussian
to the full line profile. \citet{MarDH:11} noted that such a fit
will be influenced by any excess wing emission, which will in turn
bias the computed RB asymmetry, so they fitted a single Gaussian to
the line core only. \citet{TiaMD:11} fitted a double Gaussian to
the full profile and set the LCP equal to the position of the main
component. In all cases, the data are mapped onto a finer spectral
grid (typically 10 times finer) so that the summations in equations
(\ref{eq:blue}) and (\ref{eq:red}) can be computed. Presumably a
spline interpolation is used, though details are not generally
provided in the papers. In yet another method, the LCP is taken to
be the position of peak intensity in the interpolated (higher
resolution) profile \citep{TiaMD:11}

To explore how sensitively the computed RB asymmetry depends on the
LCP, we first compare several different options (see also
\cite{TiaMD:11}). We use the central positions of single Gaussian
fits and the positions of peak intensity in spline interpolations.
Cognizant of the concerns raised by \citet{MarDH:11}, we consider
both the full line profile and subsets of the profile in the line
core. The different options are:

\begin{enumerate}
\item Single Gaussian fit to the complete spectral line profile [case 1]
\item Single Gaussian/spline fit to two blue points and one red point [case 2]
\item  Single Gaussian/spline fit to two blue and two red points [case 3]
\item  Single Gaussian/spline fit to one blue point and two red points [case 4]
\end{enumerate}

\noindent where the blue and red points in the profile are defined
with respect to the data point with maximum intensity across the
original (low resolution) profile.

\begin{table*}[htdp]
\caption{Peak intensity, line centre (reference wavelength) and the RB symmetry obtained in different
scenarios using Gaussian fitting and spline fitting.}\label{table:1}
\begin{center}
\begin{tabular}{ccccr}
\hline
{Fitting}               &           & {Peak Intensity}                                          &   {LCP }                      &{RB Asymmetry}\\
                            &           &[{erg~cm$^{-2}$~s$^{-1}$~sr$^{-1}$}] & [{\AA}]                                       &\\
\hline
Gaussian            & Case 1    &   30126.1                                                 &   202.05001               &0.00    \\
                        & Case 2    &   29850.1                                                 &   202.04917               &$-$0.02       \\
                        & Case 3    &   30124.4                                                 &   202.04956               &$-$0.01       \\
                        & Case 4    &   30124.5                                                 &   202.04956               &$-$0.01           \\
\hline
Spline              & Case 2    &   30124.5                                                 &   202.04687               &$-$0.08       \\
                        & Case 3    &   30126.4                                                 &   202.04787               &$-$0.06       \\
                        & Case 4    &   30126.3                                                 &   202.04795               &$-$0.05     \\
\hline
ICSI                    &                   & 31073.3                                                   & 202.04746              &$-$ 0.06\\
\hline
\end{tabular}
\end{center}
\end{table*}%

To determine the RB asymmetry, the steps we follow are as follows:

\begin{enumerate}

\item We determine the reference wavelength by performing either a spline or a Gaussian
fit using 4 or 5 points in
the core of the line profile or the complete line profile depending
on the scenario under consideration. The reference wavelength is
where the fit has peak intensity.

\item We map the original data onto a 25 times finer grid using a spline fit on all the points in the profile, including the wings and convert
from a wavelength scale to velocity scale assigning zero velocity to
the LCP.

\item  Then compute the RB asymmetry from $\pm$50 to $\pm$120~km~s$^{-1}$ using the method described in section~\ref{rb}.

\end{enumerate}

For our purposes here, we restrict our attention to the
\ion{Fe}{13}~$\lambda$202 line, which is considered to be one of the
cleanest spectral lines present in the EIS spectra. Throughout our
study, we use spectra that are spatially averaged over the selected
moss regions in order to improve the signal-to-noise. For this
comparison of LCP methods, we only use moss region A shown in
Figure~\ref{july1}.

Table~\ref{table:1} shows the peak intensity, line centroid and RB
asymmetry obtained by considering different scenarios for Gaussian
and spline fitting. The LCP has moved toward the red
side from case~2 to case~4 for Gaussian
fitting and from case~2 to case~4 for spline fitting case. The RB
asymmetry obtained for case~1 using Gaussian fitting is negligible.
Note that the negative (positive) values of RB asymmetry suggest red
(blue) wing enhancement. For Gaussian fitting cases~2, 3 \& 4 the
intensity in the red wing is larger then in blue wing. However, the
difference in obtained asymmetry is only about 1\%. For spline
fitting, in all three cases, there is an enhancement in the red wing
that decreases from about 8\% for case 2 to 5\% for case 4.

These results suggest that the small RB asymmetries depend very
sensitively on the LCP. We performed a simple analytical calculation
to understand this better. We took the same observed profile and
computed the RB asymmetry for many different values of the LCP
ranging between $\pm$0.25 from the position obtained using the case
3 Gaussian fit. The results show that an LCP difference of only
0.00067~{\AA} (one-thirtieth of an EIS spectral bin, or
approximately 1~km s$^{-1}$ at 200 {\AA)} produces an RB difference of
0.018. This suggests that the previously published measurements of
RB asymmetries have considerable uncertainty and should be treated
with great caution whenever the asymmetry is only a few percent.

We know that the intensity we observe in a spectral bin is the mean
intensity averaged over the bin. It is traditionally assigned to the
wavelength at the center of the bin. However, this is only
appropriate if the line profile is linear within the bin. This is
generally not the case. If the line profile is curved, then the mean
intensity should really be assigned to a position that is offset
from the bin center. Alternatively, the intensity assigned to the
center should be different from the mean intensity. \citet{KliPT:13}
have recently developed a new method called Intensity Conserving
Spline Interpolation (ICSI) that takes this into account. It is an
iterative method that, as the name implies, conserves intensity in
every spectral bin. Traditional spline fitting routines do not
conserve intensity when they force the fit to have an intensity
equal to the mean at a position midway in the bin. This can lead to
significant errors, especially in determining the LCP of the
profile.  All RB asymmetry results presented throughout the
remainder of the paper use the ICSI method to map the original data
onto a 100 times finer spectral grid. The LCP is taken to be the
position of peak intensity in the high resolution profile. The last
row in Table~\ref{table:1} provides the peak intensity, LCP and RB
asymmetry obtained using the ICSI method.

\subsection{Asymmetries in different spectral lines for two moss regions A and B}

\begin{figure}
\centering
\includegraphics[width=0.9\textwidth]{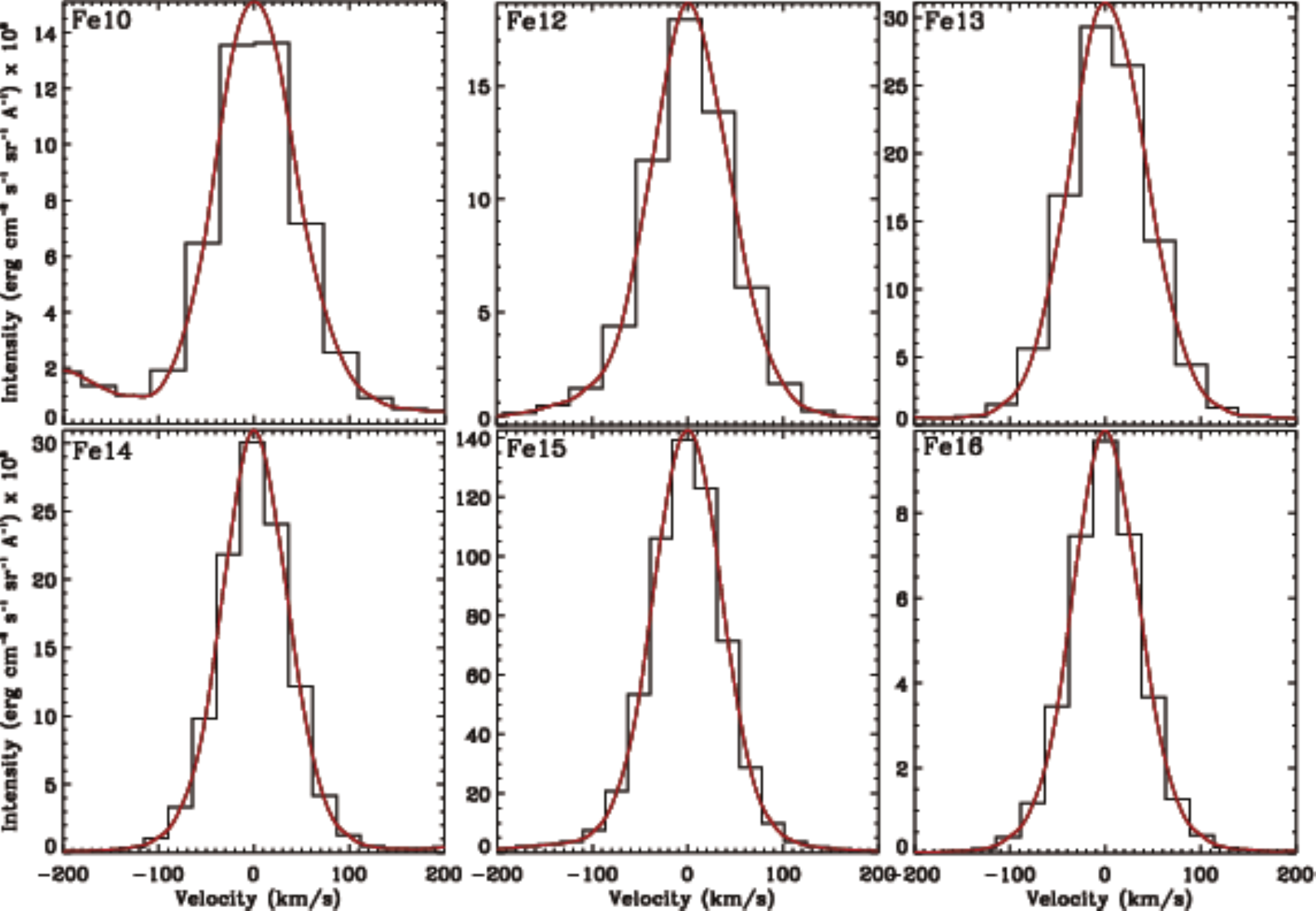}
\caption{Spectral profile from \ion{Fe}{10} to \ion{Fe}{16}. The fit is obtained by the method proposed by \cite{KliPT:13}.}\label{klimfit}
\end{figure}

We first considered moss regions A \& B in active region
\textit{AR10961} shown in Figure \ref{july1}. We measured the RB
asymmetry of \ion{Si}{7}~275.36~{\AA}, \ion{Fe}{10}~184.53~{\AA}, \ion{Fe}{12}~192.42~{\AA},
\ion{Fe}{13}~{202.04}~{\AA}, \ion{Fe}{14}~{264.79}~{\AA},
\ion{Fe}{15}~{284.16}~{\AA} and \ion{Fe}{16}~{263.98}~{\AA}.
Figure~\ref{klimfit} shows the Fe spectral lines
from moss region A over-plotted with the ICSI fits.  Notice that the
intensity integrated under the curve in each bin is equal to the
intensity of the original histogram.  This is especially important
near the peak of the profile, where the LCP is determined.

The difference of intensities between blue and red wings and
corresponding RB asymmetry (RBA) are tabulated in
Table~\ref{table:2}. Except for \ion{Si}{7} and \ion{Fe}{15}, the
intensity differences and corresponding RBAs are negative,
indicating excess red wing emission. The magnitudes of the RBAs are
generally well below 5\% and in no case exceed 10\%.  They are
largest at intermediate temperature (Fe X, XII, and XIII).

We used the Pottasch method \citep[][]{Pott:63, TriMD:10} to derive
emission measures from the intensity differences.
Figure~\ref{EM_A_B} displays the EM($T$) distributions for moss
regions A and B, where we have given the EM the same sign as the
asymmetry. The results are similar for regions A and B.
The EM has a very small value at $\log T=5.8$, becomes progressively
larger (more negative) as temperature increases, then suddenly
becomes smaller at $\log T=6.3$.  The trend of increasing magnitude
with temperature is reminiscent of what \citet{BroW:12} found,
except they detected excess blue emission rather than excess red
emission. Also, they observed a faint outflow region rather than
moss.

\begin{table*}[htdp]
\caption{Reference wavelength and RB asymmetry obtained for different spectral line for moss regions A and B using ICSI.}\label{table:2}
\begin{center}
\begin{tabular}{|l|c|r|c|r|}
\hline
{Ion}                       &\multicolumn{2}{c|}{Region A}                      &\multicolumn{2}{c}{Region B}\\
\hline
                            & I$_{B}$ {-} I$_{R}$   &   RBA         &I$_{B}$ {-} I$_{R}$        &   RBA\\
                            & [{erg~cm$^{-2}$~s$^{-1}$~sr$^{-1}$}]&     &[{erg~cm$^{-2}$~s$^{-1}$~sr$^{-1}$}]   &\\

\hline

\ion{Si}{7}                 &3.14                   &0.027              &1.43                               &0.011      \\
\ion{Fe}{10}                &{--}29.61              &{--}0.056              &{--}27.49                          &{--}0.095      \\
\ion{Fe}{12}                &{--}20.52              &{--}0.032              &{--}32.11                          &{--}0.049      \\
\ion{Fe}{13}                &{--}69.66              &{--}0.063              &{--}50.00                          &{--}0.042      \\
\ion{Fe}{14}                &{--}20.47              &{--}0.014              &{--}5.65                           &{--}0.005      \\
\ion{Fe}{15}                &69.93              &0.009              &80.99                              &0.016      \\
\ion{Fe}{16}                &{--}4.32               &{--}0.009              &{--}4.52                           &{--}0.017      \\

\hline
\end{tabular}
\end{center}
\end{table*}%

\begin{table}
\caption{Centre to limb variation of asymmetry for different spectral lines.}\label{table:3}
\centering
\begin{tabular}{|l|c|r|c|r|c|r|c|r|c|r|}
\hline
{Ion}           & \multicolumn{2}{c|}{May 01}               & \multicolumn{2}{c|}{May 02}               & \multicolumn{2}{c|}{May 03}                   & \multicolumn{2}{c|}{May 04}           & \multicolumn{2}{c|}{May 05} \\
\hline
                &I$_{B}$ {-} I$_{R}$        &   RBA         & I$_{B}$ {-} I$_{R}$       &   RBA         & I$_{B}$ {-} I$_{R}$       &   RBA             & I$_{B}$ {-} I$_{R}$       &   RBA         & I$_{B}$ {-} I$_{R}$       &   RBA\\
\hline
\ion{Si}{7}     &2.52                           & 0.021         &2.37                           &0.022          &0.10                           &0.001              &1.00                           &0.008      &1.00                           &0.007\\
\ion{Fe}{12}    &{-}10.65                       &{--}0.020          &4.46                           &0.008          &13.89                      &0.022              &{-}20.60                       &-0.031     &29.69                      &0.045\\
\ion{Fe}{13}    &{-}33.56                       &{-}0.042           &{-}22.08                       &{-}0.027           &12.38                      &0.012              &{-}2.86                        &{-}0.002       &{-}1.88                        &{-}0.001\\
\ion{Fe}{14}    &4.14                           &0.003          &7.01                           &0.005          &68.01                      &0.036              &{-}47.52                       &{-}0.024       &27.26                      &0.018\\
\ion{Fe}{15}    &{-}76.53                       &{-}0.012           &186.11                     &0.034          &{-}91.86                       &{-}0.009               &278.79                     &0.033      &140.63                     &0.019\\
\ion{Fe}{16}    &3.58                           &0.008          &5.70                           &0.012          &16.30                      &0.022              &{-}15.77                       &{-}0.021       &1.67                           &0.004\\
\hline
\end{tabular}
\end{table}%
\subsection{Center to limb variation of asymmetries in different spectral lines}

We next studied the center-to-limb variation of the RBA for the moss
region in active region \textit{AR10953} shown in Figure \ref{may1}.
Center-to-limb variations are important for understanding time evolution and
geometry (line of sight) effects. For example, the center-to-limb
behavior of the RBA in the faint outflow regions studied by
\citet{BryYD:10} and \citet{TiaMX:12} indicate high-speed upflows
inclined outward from the center of the active region.  This is the
expected tilt of the magnetic field at the periphery of the active
region where the outflow regions are located.  As pointed out by
\citet{TiaMX:12}, this is also strong evidence that the RBAs are
not artifacts of blends.  Blends would have an equivalent effect at
all disk locations, not preferentially at disk center or the east
limb or the west limb.

Table~\ref{table:3} shows the intensity differences and RBAs tracked
over 5 days. The EM($T$) distributions are plotted in Figure~\ref{EM_ALL}.
A combination of positive (enhanced blue wing) and
negative (enhanced red wind) values are present. The magnitudes of
the asymmetries are small. The largest is 4.5\%, and most are $<
2\%$. The measurements have a rather random appearance, with no
obvious patterns, either in temperature or in time. The only
consistency is that the EM is very small at $\log T=5.8$ on all of
the days.

\section{Summary and Discussion} \label{conc}

\begin{figure}
\centering
\includegraphics[width=0.8\textwidth]{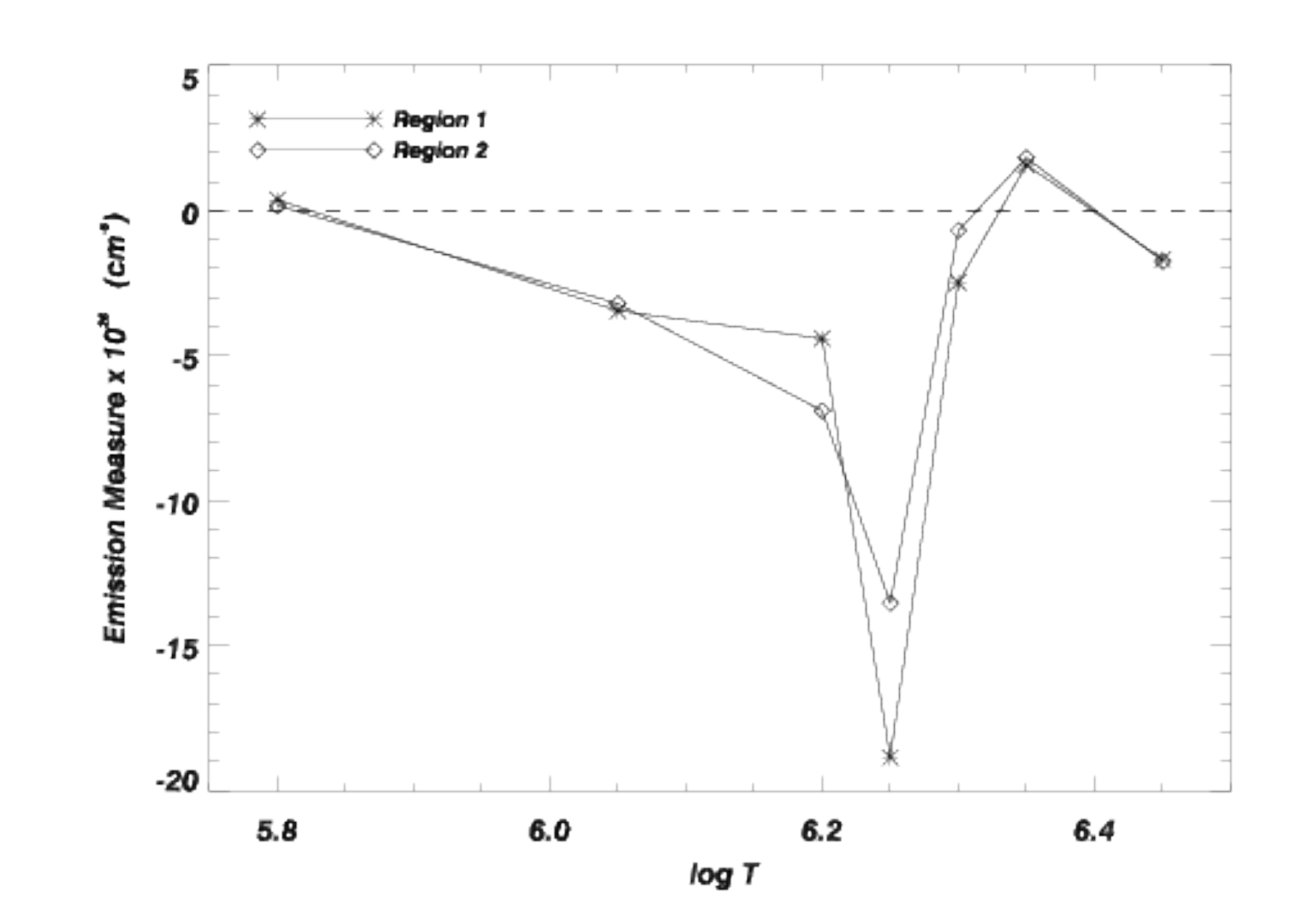}
\caption{Emission measure distribution obtained using the difference in intensities in moss A and B.}\label{EM_A_B}
\end{figure}

\begin{figure}
\centering
\includegraphics[width=0.8\textwidth]{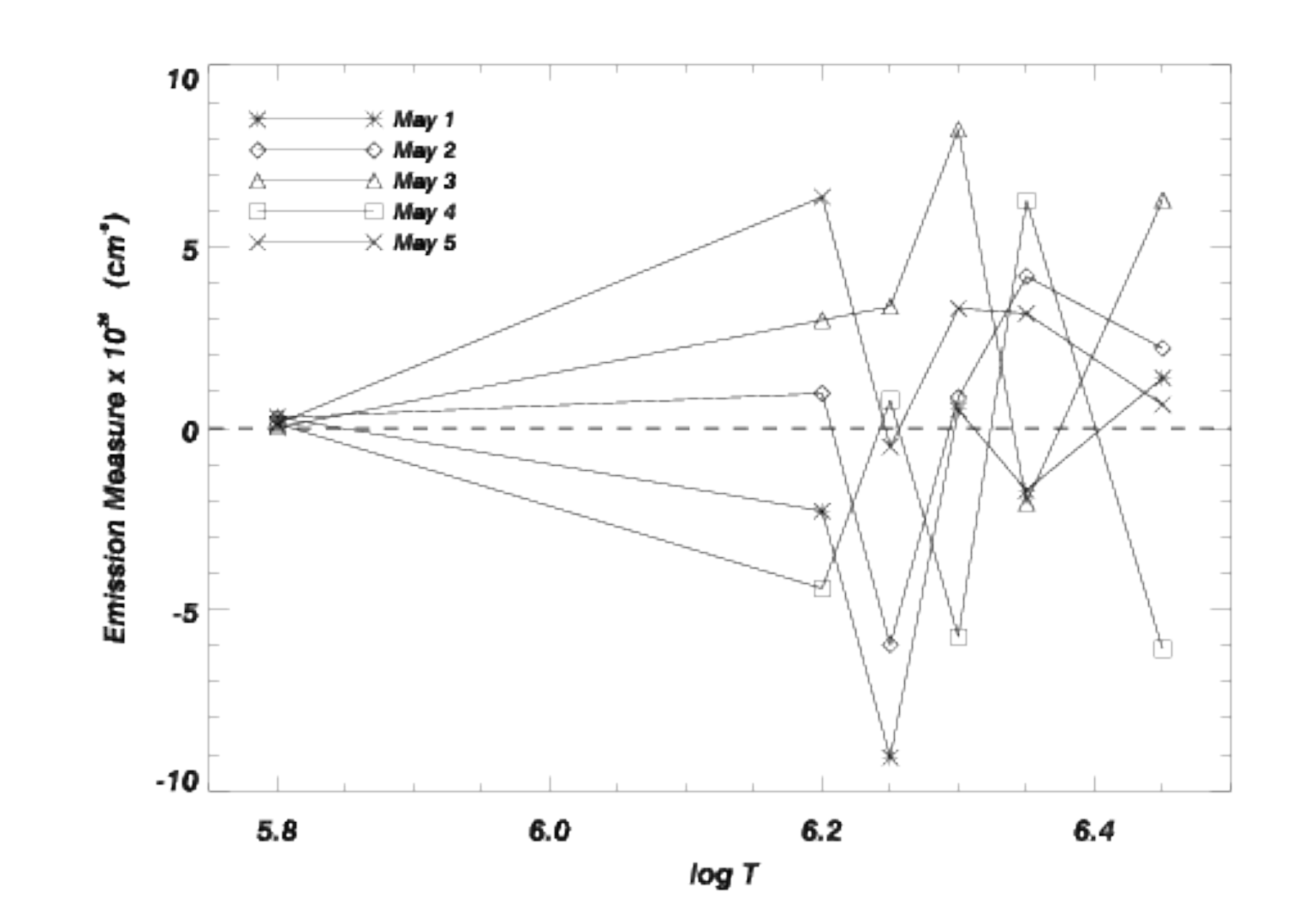}
\caption{Emission measure distribution obtained using the difference in intensities in moss region tracked for five
consecutive days.}\label{EM_ALL}
\end{figure}

The measurements presented in Tables \ref{table:2} and \ref{table:3}
and Figures \ref{EM_A_B} and \ref{EM_ALL} paint a confusing picture.
The situation is made worse if we combine EM distributions from the
two figures in a single plot. We are forced to conclude that the RBA
and EM measurements are likely dominated by uncertainties. This
should not be surprising, given the discussion in Section 4 and the
fact that the measured RBA values are very small.  A 2\% RBA is
equivalent to an error in the line center position (LCP) of less
than one-thirtieth of an EIS spectral bin.

The measurements are nonetheless extremely useful for placing upper
limits on the emission measure of any rapidly flowing material that
would emit in the line wings.  The most consistent aspect of our
measurements is the EM at $\log T=5.8$. At all moss locations and on
all days, EM $< 4\times10^{25}$ cm$^{-5}$ at this lowest temperature. At the
other temperatures, EM is mostly $<5\times10^{26}$ cm$^{-5}$. However, since the sign of the asymmetry
varies in seemingly random ways, we suggest that the true EM is
actually smaller and masked by errors in the measurements.

We can compare these upper limits with the EM we measured previously
in the inter-moss regions. For \textit{AR10961} the inter-moss EM
peaks at about $10^{28}$ cm$^{-5}$, and for \textit{AR10953} it
peaks at about $3\times10^{28}$ cm$^{-5}$ \citep{TriKM:11}. These
inter-moss measurements represent the coronal plasma that is
magnetically linked to the plasma at the moss footpoints.  We can
therefore compare the emission measures in the same way that
\citet{Kli:12} did for blue wing and line core emission of
individual hot coronal lines. We find that the EM of the rapidly
upflowing plasma is approximately two orders of magnitude less than
the EM of the stationary (or slowly downflowing) plasma contained
within the same magnetic loops.  In stark contrast, Klimchuk argued
that the EM of the upflow would need to exceed the EM of the
downflow if type II spicules are the primary source of coronal
plasma.

In summary, \citet{Kli:12} showed that 1. spicules do not provide
enough pre-heated plasma to fill the corona, and 2. even if they
did, additional heating would be required to keep the plasma hot as
it expands upward. The question remained as to whether spicules can
help explain the corona by injecting plasma at cooler temperatures
($< 2$ MK), which then gets heated to coronal values at higher
altitudes. Our study indicates that this is not the case in active
regions, at least for injection temperatures $> 0.6$ MK ($\log T
> 5.8$). Spicules carry a large mass flux at much
cooler temperatures ($< 0.1$ MK), but most of this mass falls back
to the surface in cool state after reaching a maximum height of
about $10^4$ km.  We are aware of no evidence for a large upward
mass flux in the range $0.1 < T < 0.6$ MK, but further
investigation is warranted.

We end by noting that these conclusions apply not only to spicules,
but to any process that suddenly heats and accelerates chromospheric
plasma (e.g., a chromospheric nanoflare).  We conclude that the
traditional picture of coronal heating and chromospheric evaporation
remains the most likely explanation of the corona.

\acknowledgments{Hinode is a Japanese mission developed and launched
by ISAS/JAXA, collaborating with NAOJ as a domestic partner, NASA
and STFC (UK) as international partners. Scientific operation of the
Hinode mission is conducted by the Hinode science team organized at
ISAS/JAXA. This team mainly consists of scientists from institutes
in the partner countries. Support for the post-launch operation is
provided by JAXA and NAOJ (Japan), STFC (U.K.), NASA, ESA, and NSC
(Norway). CHIANTI is a collaborative project involving researchers
at NRL (USA) RAL (UK), and the Universities of Cambridge (UK),
George Mason (USA), and Florence (Italy).  The work of JAK was
supported by the NASA Supporting Research and Technology Program.
The authors benefited from participation in the International Space
Science Institute team led by S. Bradshaw and H. Mason. The authors also
thank H. Mason, P. R. Young and Srividya Subramanian for various discussions.}

\bibliography{reference}
\clearpage




\end{document}